\documentclass[twocolumn]{revtex4}

\usepackage{amsmath,amssymb}
\usepackage{graphicx}
\usepackage{dcolumn}
\usepackage{bm}

\def\PRL{Phys. Rev. Lett. }
\def\RMP{Rev. Mod. Phys. }
\def\EPJ{Eur. Phys. Journ. B }
\def\SCI{Science }
\def\PHC{Physica C }
\def\JETP{Sov. Phys. JETP }

\def\PRB{Phys. Rev. B }

\def\be{\begin{equation}}
\def\ee{\end{equation}}
\def\bea{\begin{eqnarray}}
\def\eea{\end{eqnarray}}

\begin{document}

\title{Time Reversal Breaking Superconducting State\\ in
the Phase Diagram of the Cuprates}

\author{G. Sangiovanni, M. Capone, S. Caprara}
\affiliation{Dipartimento di Fisica, Universit\`a di Roma ``La Sapienza'',
and  \\INFM Center for Statistical Mechanics and Complexity,
Piazzale Aldo Moro 2, I-00185 Roma, Italy}
\email{giorgio.sangiovanni@roma1.infn.it}

\begin{abstract}
We review and extend a previous study on the symmetry of the superconducting state, stimulated by recent tunneling and Andreev reflection measurements giving robust evidences for the existence of a $d_{x^2-y^2}+id_{xy}$ order parameter in the overdoped regime of two different cuprates. Looking for a possible second-order phase transition from a standard $d_{x^2-y^2}$ to a mixed and time reversal breaking state, we confirm the results of our previous analysis on La$_{2-x}$Sr$_{x}$CuO$_{4}$. In the case of Y$_{1-y}$Ca$_y$Ba$_2$Cu$_3$O$_{7-x}$ as well, among all the allowed symmetries, the $d_{x^2-y^2}+id_{xy}$ is the most favored one and the unconventional state is likely to occur in a small dome of the phase diagram located in the optimal-overdoped region and at very low temperatures.
\end{abstract}

\maketitle

\section{Introduction} \label{sec:intro}

Recent tunneling and Andreev reflection experiments\cite{deutscher,gonnelli}
on high-$T_c$ cuprate superconductors (HTCS) have shown small deviations from the pure $d_{x^2 - y^2}$ symmetry of the superconducting gap\cite{tsueirev}.
Deutscher and co-workers have determined the doping dependence of the zero-bias conductance peak in tunneling experiments on (110) oriented Y$_{1-y}$Ca$_y$Ba$_2$Cu$_3$O$_{7-x}$ (YBCO) films and have shown that this peak, which is associated to the $d_{x^2 - y^2}$ symmetry\cite{hu}, is spontaneously splitted in some of the measured samples signaling the presence of a small secondary component of the bulk superconducting gap\cite{fogel}. This secondary component, which removes the nodal lines of the $d_{x^2 - y^2}$ primary one along the $\Gamma{\mathrm X}$ (diagonal) directions of the Brillouin zone, is compatible with a time-reversal breaking $id_{xy}$ gap function.
The splitting of the zero bias conductance peak is non-zero only for $\delta > \delta_{opt}$ (slightly overdoped samples), and vanishes for $\delta < \delta_{opt}$ (underdoped samples). Heavily overdoped YBCO samples are not available.
This scenario has been subsequently confirmed by Gonnelli and co-workers who performed a re-analysis of their Andreev reflection data on La$_{2-x}$Sr$_{x}$CuO$_{4}$ (LSCO)\cite{gonnelli} originally interpreted with a $d_{x^2 - y^2}+s$ symmetry. They showed that a better agreement with the experimental results is obtained assuming a $d_{x^2 - y^2}+id_{xy}$ pairing symmetry\cite{gonnprivate} and found that the amplitude of the splitting is not monotonic with doping. They are able to perform measurements on heavily overdoped LSCO samples and find the sign for a closure of the $d_{x^2 - y^2}+id_{xy}$ region in the overdoped.

In the past Balatsky\cite{balatsky} has
studied the properties of the magnetic-field-induced transition to the
$d_{x^2 - y^2}+id_{xy}$ state proposed by Laughlin\cite{laughlin} to
explain the thermal conductivity plateaux in Bi$_2$Sr$_2$CaCu$_2$O$_8$\cite{krishana}. More recently a quantum critical point separating a $d_{x^2 - y^2}+id_{xy}$ from a pure $d_{x^2 - y^2}$ phase has been suggested\cite{vojta}.
Recently we have given, in collaboration with other authors, a simple analysis of the gap equation based on general symmetry arguments that indicates the existence of a $d_{x^2 - y^2}+id_{xy}$ superconducting region in the phase diagram of the cuprates\cite{sangiova}. In accordance with the recent experimental results\cite{deutscher,gonnelli} we have found that the unconventional pairing state should be stable in the optimal-overdoped regime and at very low temperatures.

\section{Symmetry analysis and BCS equation} \label{sec:BCS}

In order to describe the transition between two different superconducting phases, as the one from a pure $d_{x^2 - y^2}$ to a mixed $d_{x^2 - y^2}+id_{xy}$ state, we can use the classification of the superconducting classes based on the symmetry properties under crystal point transformations\cite{volovik85}.

We start by imposing two constraints on the order parameter:
\begin{description}
\item[{\it i}.] $|\Delta_{g{\bf k}}| = |\Delta_{\bf k}|$ where $g{\bf k}$ denotes the vector obtained from ${\bf k}$ after the action of one of the $N_G$ group operators $O_g$.
\item[{\it ii}.] $\Delta_{{\bf k}+{\bf K}} = \Delta_{\bf k}$ where ${\bf K}$ is a reciprocal lattice vector
\end{description}
The first condition assures that the dispersion $E_{\bf k}=\sqrt{\xi_{\bf k}^2 + |\Delta_{\bf k}^2|}$ that enters the physical observables is a total symmetric function with respect to the lattice symmetry\cite{notasimm}, while the second is simply to preserve the crystal periodicity.
Given the strong evidence of singlet pairing in HTCS we require that
\begin{description}
\item[{\it iii}.] $\Delta_{-{\bf k}} = \Delta_{\bf k}$
\end{description}
This further requirement imposes us to work with even functions of ${\bf k}$, because of the antisymmetric spin part of the wave function.

Due to the strong anisotropy of the copper oxides we take a two-dimensional square lattice described by the non abelian $\mathcal{C}_{4v}$ point group.
\begin{table}[htbp]
\begin{center}
\begin{tabular}{c|*{5}{r}|l}
${\mathcal C}_{4v}$ & $E$ & $C_4^2$ &  $C_4(2)$ & $\sigma_v(2)$ & $\sigma_{v'}(2)$ &\\
\hline
      ${\mbox A}_1 $         & 1 & 1       &  1        & 1             & 1       & $s^+$         \\
      ${\mbox A}_2 $         & 1 & 1       &  1        & -1            & -1      & $s^-$          \\
      ${\mbox B}_1 $         & 1 & 1       &  -1       & 1             & -1      & $d^+$           \\
      ${\mbox B}_2 $         & 1 & 1       &  -1       & -1            & 1       & $d^-$          \\
      ${\mbox E} $           & 2 & -2      &  0        & 0             & 0       & $e$          \\
\end{tabular}
\end{center}
\caption{$\mathcal{C}_{4v}$ characters table. In the last column on the right we have labeled the symmetry of the base functions for each representation.}
\label{tabellaC4v}
\end{table}
This group has eight elements ($N_G=8$): the identity operator $E$, one $2$-fold rotation $C_4^2$ and two $4$-fold rotations $C_4$ about the $Z$ axis perpendicular to the $XY$ lattice plane and four reflections $\sigma_v$ and $\sigma_{v'}$ in the four vertical planes.

We can write the order parameter as
\begin{equation} \label{gap1}
\Delta_{\bf k}=\sum_{\mu} \Delta^{(\mu)}_{\bf k}
\end{equation}
where the sum over $\mu$ runs over all the irreducible representations.
Using standard methods of group theory\cite{ostlund} one can determine the expressions for the various $\Delta^{(\mu)}_{\bf k}$ in (\ref{gap1}):
\begin{equation} \label{gap2}
\Delta^{(\mu)}_{\bf k}={\sum_{\bf r}}^\prime \Delta_{\bf r}^{(\mu)} w_{\bf r}^{(\mu)}({\bf k})
\end{equation}
where the primed sum is restricted to non-equivalent sites of the lattice, that is sites ${\bf r}=(x,y)$ that cannot be obtained by means of the action of any of the $O_g$ operators on another site ${\bf r'}$.
We deal, in this way, only with gap amplitudes on a particular lattice distance while the sign relation between all the equivalent sites is entirely accounted for by the harmonics $w_{\bf r}^{(\mu)}({\bf k})$
\be \label{wharm}
\left\{
\begin{array}{rcl}
w_{\bf r}^{s^+}({\bf k}) & = & \cos k_x x \cos k_y y + \cos k_x y \cos k_y x  \\
\\
w_{\bf r}^{s^-}({\bf k}) & = & - \sin k_x x \sin k_y y + \sin k_x y \sin k_y x \\
\\
w_{\bf r}^{d^+}({\bf k}) & = & \cos k_x x \cos k_y y - \cos k_x y \cos k_y x  \\
\\
w_{\bf r}^{d^-}({\bf k}) & = & - \sin k_x x \sin k_y y - \sin k_x y \sin k_y x  \\
\end{array}
\right.
\end{equation}
For ${\bf r} = 0$, the only non-zero harmonic belongs to the  $s^+$ representation, and corresponds to the isotropic $s$-wave. For ${\bf r}=(\pm 1,0);(0, \pm 1)\equiv {\bf 1}$ the $s^+$ and $d^+$ representations are associated with the extended $s_{x^2+y^2}$ and $d_{x^2 - y^2}$ waves, respectively. For ${\bf r}= (\pm 1,\pm 1)\equiv {\bf 2}$, the non-vanishing harmonics are related to the $s_{xy}$ and $d_{xy}$ waves ($s^+$ and $d^-$ representations).
In (\ref{wharm}) we have not included the basis functions for the two-dimensional representation ${\mbox E}$ because requirement {{\it iii}.} rules out odd functions of ${\bf k}$. This is a peculiarity of the bi-dimensional square lattice.
The $\mathcal{D}_{4h}$ group for instance, proper to a system in which the coupling between the different $Cu$-$O$ planes is not negligible, possesses a two dimensional representation which is even with respect to inversion and in that case it must be considered.


It can be seen that, using the expansion defined by Eq.(\ref{gap1}) and (\ref{gap2}), the BCS equation becomes
\begin{equation} \label{BCSeq}
\Delta_{\bf r}^{(\mu)} = - \frac{2}{\lambda_{\bf r}} \hat{V}({\bf r}) {\sum_{\bf r'}}^\prime \Delta_{\bf r'}^{(\mu)}
\sum_{\bf k'} w_{\bf r}^{(\mu)}({\bf k'}) w_{\bf r'}^{(\mu)}({\bf k'}) K({\bf k'})
\end{equation}
where the kernel $K({\bf k})$ is defined as
\begin{equation} \label{kernelK}
K({\bf k}) = \frac {\mbox{tanh} \frac{\beta}{2} \sqrt{\xi^2_{\bf k}+\left|\sum_{\nu} {\sum}'_{\tilde{\bf r}} {\Delta_{\tilde{\bf r}}^{(\nu)}} w_{\tilde{\bf r}}^{(\nu)}({\bf k})\right|^2}}
{2\sqrt{\xi^2_{\bf k}+\left|\sum_{\nu} {\sum}'_{\tilde{\bf r}} {\Delta_{\tilde{\bf r}}^{(\nu)}} w_{\tilde{\bf r}}^{(\nu)}({\bf k})\right|^2}}
\end{equation}
and $\lambda_{\bf r}$ is a multiplicity factor equal to $1$ for a generic ${\bf r}$, $2$ for a site on the axis or on the bisectors of the lattice and $8$ for ${\bf r}=(0,0)$.
It must be noted that the form (\ref{BCSeq}) represents the mean-field equation for a system described by a generic static pairwise interaction between the electrons $V_{{\bf k}{\bf k'}}$ (with $\hat{V}({\bf r})$ the corresponding real space Fourier transform) and it is obtained without any previous decoupling of the interaction term in a given symmetry channel. It consists of an infinite dimension homogeneous problem which, in general, will be of difficult solution. Nevertheless it can be observed that the appearance of a harmonic labeled by ${\bf r}$ is always  controlled by the corresponding $\hat{V}({\bf r})$. If the attractive potential is therefore a rapidly decreasing function of ${\bf r}$ we can restrict ourselves to the smallest distances bonds.
We introduce the short-hand notations $V_0,V_1,V_2,\ldots$ for the
on-site, nearest-neighbor (nn), next-to-nearest-neighbor (nnn), \ldots
components of $\hat{V}({\bf r})$. Due to the repulsive character of the local interaction $V_0$ in the HTCS, the isotropic $s$-wave is ruled out. We consider here attractive (negative) nn ($V_1$) and nnn ($V_2$) interactions. For the following discussion we do not take into account longer range terms.

We have not yet made use explicitly of the first requirement on the invariance of the gap amplitude $|\Delta_{\bf k}|$ under point group transformations. Its effect is to force the gap function to transform either like a single representation, or like a complex (time-reversal breaking) combination of the form $\Delta^\eta_{\bf k} + i \Delta^\zeta_{\bf k}$, with $\Delta^\eta_{\bf k}$ and $\Delta^\zeta_{\bf k}$ belonging to two different representations of the point group.
To describe this latter possibility, in the case of $V_0>0$, $V_1<0$, $V_2<0$, we take the complex combination $\Delta_{\eta} w^{\eta}_{\bf 1}({\bf k})+i\Delta_{\zeta} w^{\zeta}_{\bf 2}({\bf k})$, with both $\Delta_{\eta}$ and $\Delta_{\zeta}$ real. Each gap parameter affects the other only through the quasi-particle spectrum and the BCS equations read
\begin{eqnarray}
\frac{1}{|V_1|}&=&
\int \frac{d{\bf k}}{4\pi^2}
\left(w^{\eta}_{\bf 1}({\bf k})\right)^2 
\frac{\tanh (\beta E_{\bf k}/2)}{2 E_{\bf k}} 
\label{BCS1}\\
\frac{1}{|V_2|}&=&
\int \frac{d{\bf k}}{4\pi^2}
\left(w^{\zeta}_{\bf 2}({\bf k})\right)^2 
\frac{\tanh (\beta E_{\bf k}/2)}{2 E_{\bf k}}
\label{BCS2}
\end{eqnarray}
where $\beta$ is the inverse temperature and
$$E_{\bf k}=\left[\xi_{\bf k}^2+\left(\Delta_{\eta} w^{\eta}_{\bf 1}({\bf k})\right)^2+\left(\Delta_{\zeta} w^{\zeta}_{\bf 2}({\bf k})\right)^2\right]^{1/2}$$
with $\eta=s^+,d^+$, $\zeta=s^+,d^-$ ($\eta \ne \zeta$).
For typical values of the particle density in the HTCS, the Van Hove singularity (VHS) of the two-dimensional DOS strongly favors the $d_{x^2 - y^2}$ symmetry, which thus represents the principal component of the superconducting gap, in agreement with the phenomenology\cite{marki}.
We take henceforth $\eta=d^+$ in Eqs. (\ref{BCS1}),(\ref{BCS2}).

As we mentioned above the spectrum for the $\zeta$ component is
(pseudo)gapped by the principal $d_{x^2-y^2}$ gap.
As a consequence a critical value $|V_2^{cr}|$ for the onset of
a mixed order parameter exists, as opposed to the case of the Cooper
instability in the metallic phase.
The critical coupling for both $d_{xy}$ and $s_{xy}$ secondary pairing
is determined by  solving the BCS equations (\ref{BCS1}),(\ref{BCS2})
with $\Delta_{\zeta}=0$ and $\zeta=d^-,s^+$ respectively\cite{sangiova}.
Numerical solutions show that the $d_{x^2-y^2}+id_{xy}$ symmetry is much favored with respect to the $d_{x^2-y^2}+is_{xy}$ as one expects from a qualitative analysis of Eq.(\ref{BCS1}),(\ref{BCS2}).
Moreover, the continuous transition from $d_{x^2-y^2}$ to $d_{x^2-y^2}+is_{xy}$ is generally prevented by a first-order transition from pure $d_{x^2-y^2}$ to pure $s_{xy}$, due to the strong competition between the two harmonics, which makes mixing unfavorable. This scenario does not find experimental support.
The continuous transition from $d_{x^2 - y^2}$ to $d_{x^2 - y^2}+id_{xy}$ gap is therefore the most natural candidate if a pure $d_{x^2 - y^2}$ is to be modified by a small secondary component, as suggested by the experiments\cite{deutscher,gonnelli}.
As far as the dependence of the $d_{xy}$ critical coupling on the value of $|V_1|$ is concerned it can be realized that the more $|V_1|$ grows the smaller is the ratio $|V_2^{cr}|/|V_1|$ so as to favor the transition to the $d_{x^2-y^2}+id_{xy}$ state in the intermediate-to-strong coupling regime.

\section{Results and Conclusions} \label{sec:results}

\begin{figure}[htbp]
\begin{center}
\includegraphics[width=8cm]{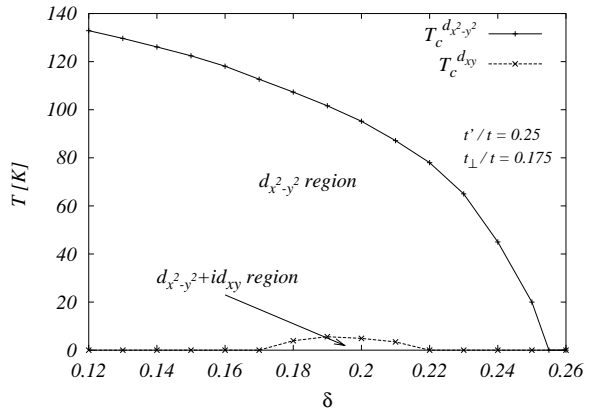}
\end{center}
\caption{hase diagram for $d_{x^2 - y^2}$ and $d_{x^2-y^2} +
id_{xy}$ pairing in the HTCS. The parameters are appropriate for
YBCO.} \label{phd}
\end{figure}

In this section we show numerical solutions of Eq.(\ref{BCS1}),(\ref{BCS2}) obtained studying the dependence on hole concentration ($\delta$) and on band parameters.
In a previous work\cite{sangiova} the phase diagram for the $d_{x^2-y^2}+id_{xy}$ superconducting region for the case of LSCO (closed Fermi surface) was derived. Here we want to show the case of bilayer compounds like Bi$_2$Sr$_2$CaCu$_2$O$_{8+\delta}$ (Bi2212) and YBCO (open Fermi surface).
Our rigid band dispersion is thus $\xi^{\pm}_{\bf k}=-2t(\cos k_x+\cos k_y)+4t^\prime\cos k_x \cos k_y \pm \frac{1}{2}t_\perp (\cos k_x-\cos k_y )^2-\mu$ where $t$ and $t'$ are nn and nnn hoppings and $\mu$ is the chemical potential. We take $t=200$ m$eV$, $t^\prime=50$ m$eV$ and $t_\perp=35$ m$eV$.
For the case of Bi2212, ARPES measurements show a value for $2t_\perp$\cite{bilayer} that varies from 88 m$eV$ to 140 m$eV$. Unfortunately we are not aware of similar data on YBCO, but the bare values of $t_\perp$ in Bi2212 and YBCO are similar, so that we can expect similar renormalized values.
In addition we introduce a cut-off $\omega_0$ in the integrals which measures the typical energy scale of the attractive potential\cite{notacutoff}.
As shown in Fig. \ref{phd}, for each doping $\delta$, we fix $V_1$
such that the critical temperature for $d_{x^2-y^2}$ pairing
coincides with the pseudogap temperature $T^*$   (which we
interpret as the temperature for Cooper pair formation without
phase coherence in the underdoped regime), and smoothly connects
to $T_c$ around optimal doping. Then we fix a value for the ratio
$V_2/V_1$ that never exceeds the physical threshold of $1$. The
connection between this value and the presence of a quantum
critical point located near optimum doping is given a previous
work\cite{sangiova}.

We find that the mixed $d_{x^2 - y^2}+id_{xy}$ order parameter is then stabilized in a small dome slightly shifted in the overdoped region, $0.17 < \delta < 0.22$. The $d_{x^2 - y^2}+id_{xy}$ transition temperature is always only a small fraction of the $d_{x^2-y^2}$ one and the same ratio is obtained between the gap amplitudes.
In the extreme overdoped region the system approaches weak coupling determining the closure of the dome. On the other hand the vanishing of the $d_{xy}$ component in the underdoped region is due to band structure effects.

We finally summarize the results of this analysis:
{\\1.} As found in our previous work\cite{sangiova} a mixed order parameter of the form $d_{x^2-y^2} +id_{xy}$ can be stabilized in a small dome contained in the larger $d_{x^2-y^2}$ region.
The $d_{xy}$ component does not grow indefinitely with increasing doping
and it is always smaller than the $d_{x^2-y^2}$.
A recent re-analysis of the LSCO data with a $d_{x^2-y^2} +id_{xy}$ gap
parameter\cite{gonnprivate}, clearly displays the closure of the dome in the overdoped.
The confirmation of our prediction in YBCO calls instead for further measurements at larger doping in the overdoped region.
{\\2.} The intermediate-to-strong coupling
regime is a necessary condition for the existence of the mixed order
parameter.
{\\3.} $d_{x^2-y^2} +id_{xy}$ pairing
is more likely to occur in closed Fermi surface materials
such as LSCO or electron doped materials, and in compounds where an
interlayer hopping splits the VHS.



\section*{Acknowledgments}
We thank Claudio Castellani, Carlo Di Castro and Marco Grilli for their collaboration.



\begin{thebibliography}{99}
\bibitem{notasimm}
We want to describe here phases that do not break the crystal symmetry.
\end{thebibliography}
\end{document}